\documentclass[11pt]{article}
\usepackage[english]{babel}
\usepackage{graphicx}
\usepackage{xcolor}
\usepackage{cite}

\usepackage{amsmath}
\usepackage{amssymb}
\usepackage{authblk}
\usepackage{lineno}

\usepackage{color}
\usepackage[normalem]{ulem}
\definecolor{bbcolor}{rgb}{0,0.1,0.7}
\definecolor{ascolor}{rgb}{1,0,1}

\DeclareRobustCommand\bbout{\bgroup\markoverwith{\color{bbcolor}{\rule[0.4ex]{2pt}{0.8pt}}}\ULon}
\DeclareRobustCommand\asout{\bgroup\markoverwith{\color{ascolor}{\rule[0.4ex]{2pt}{0.8pt}}}\ULon}

\newcommand{\tdm}[1]{\mbox{\boldmath$#1$}}

\def\lapproxeq{\lower .7ex\hbox{$\;\stackrel{\textstyle
<}{\sim}\;$}}
\def\gapproxeq{\lower .7ex\hbox{$\;\stackrel{\textstyle
>}{\sim}\;$}}

\title{\bf Longitudinal structure function $F_L$ at low $Q^2$ and low~$x$ with model for higher twist: an update}
\author[1]{Barbara Bade\l ek}
\author[2]{ Anna M. Sta\'sto}
\affil[1]{\small \it Faculty of Physics,
University of Warsaw, 02-093 Warsaw, Poland}
\affil[2]{\small \it Department of Physics, Penn State University, University Park, PA 16802, USA}

\begin{document}
\maketitle
\begin{abstract}
A reanalysis of  the model for the longitudinal 
structure function $F_L (x,Q^2)$ at low~$x$ and low $Q^2$ was
undertaken, in view of the advent of the  EIC. 
The model is based on the photon-gluon
fusion mechanism suitably extrapolated to the region of low $Q^2$. It includes
the kinematic constraint
$F_L\sim Q^4$ as $Q^2\rightarrow$ 0 and   higher twist  contribution which vanishes as $Q^2 \rightarrow \infty$.
 Revised model was critically updated 
and compared to the presently available data.
\end{abstract}

\section{Introduction}
\label{section_introduction}

Knowledge of both $F_2$ and $F_L$ structure functions,
from the photoproduction to the deep inelastic
region is needed in the calculations of QED radiative
corrections 
 to the data from the Deep Inelastic Scattering (DIS) process, l+p $\rightarrow$ l$^\prime$+X (l, l$^\prime$ are leptons), Ref.  \cite{Badelek:1994uq}.
This knowledge is also essential in  verifying the sum rules, e.g. the Gottfried sum rule or (in case of the spin-dependent function $g_1$) the Bjorken sum rule \cite{COMPASS:2015mhb}, a fundamental relation 
of Quantum ChromoDynamics (QCD). Thus  the electroproduction structure functions,
$F_2$,  $F_L$ and  $g_1$ in the full kinematic region are indispensable in the data analysis, especially 
in the context of the future DIS facilities 
like the Electron Ion Collider  (EIC), currently planned in the US \cite{AbdulKhalek:2021gbh}. In this paper we shall consider the spin-independent longitudinal structure function $F_L$.

The structure functions depend on two variables: $x$ and $Q^2$,
 conventionally defined as 
$x=Q^2/(2p\cdot q)$ and $Q^2=-q^2$ where $q$ and $p$ denote the four momentum 
transfer between the incident and scattered leptons and the four momentum 
of the target proton respectively; these four-vectors and the four-momentum $p_l$ of the incident lepton, define the inelasticity $y$ as $y=p_l\cdot p/(q\cdot p)$. 
Unlike $F_2$ and  $g_1$, the experimental data for $F_L$ are rather scarce 
for the low $Q^2$ values and thus an extrapolation to this region 
needs to be performed with a physically motivated model with least number 
of free parameters. The model should also contain an extrapolation 
to the region of low $x$, as data both from fixed-target and 
 colliders correlate these two kinematic regions. 
Such model for $F_L$ was proposed some time ago 
 \cite{Badelek:1996ap}. It was based on the $k_T$-factorization 
formula which involves  the unintegrated 
gluon density with their transverse momentum ($k_T$) dependence. 
The $k_T$-factorization was derived in the high-energy limit of $s \gg |t|$, where $s$ is the centre-of-mass energy squared in the scattering process and $t$ is the four-momentum transfer in the $t$-channel. It was derived for processes like heavy quark production in hadron-hadron collisions as well as for DIS.
In principle the unintegrated gluon distribution should be obtained from the Balitsky-Fadin-Kuraev-Lipatov (BFKL) equation, which 
 resums  the large logarithms of energy  (or small  $x$).
In the model used in Ref.\cite{Badelek:1996ap} the unintegrated gluon distribution function was constructed from the collinear 
integrated gluon density through the logarithmic derivative over the scale dependence. 
This effectively neglected higher order small $x$ contributions in the gluon anomalous dimension; these contributions are expected to be significant only at extremely low values of $x$. 

The $k_T$-factorization formula  was then extrapolated to the low $Q^2$ region by  
introducing the cutoff on the low quark transverse momenta.  
This region is dominated by the soft physics with
a higher twist contribution to $F_L$ vanishing 
at large $Q^2$. 
In the model it was treated phenomenologically and its normalization 
was determined from the (non-perturbative) part
of the structure function $F_2$. The model embodied 
the kinematic constraint $F_L \sim Q^4$ in 
the limit $Q^2\rightarrow 0$ for fixed $2p\cdot q$. 
It thus contained only physically motivated  parameters.

In this paper we revisit the model of Ref. \cite{Badelek:1996ap} 
for the extrapolation of $F_L(x,Q^2)$ to the region 
of low values of $Q^2$ at low $x$. We have included the updated
parametrizations of the Parton Distribution Functions (PDFs) 
for the quark and gluon densities in the proton, checked the sensitivity of the results to the assumed quark masses and tested the gluon distributions supplemented with the Sudakov form-factor.

The calculations have been compared with 
high $Q^2$ measurements of $F_L$ by HERA 
and with
low $Q^2$ data by SLAC and JLab. 
Here one has to be aware that most of the SLAC and JLab data are in the regime
which only marginally overlaps with 
the region of applicability of the model, 
as they correspond to  rather high values of 
$x \gapproxeq 0.1$\ . 

The outline of the paper is as follows. In the next section we discuss the general properties of the longitudinal structure function, in Sec.~\ref{section_kT} the $k_T$-factorization is introduced, while in Sec.~\ref{section_model} the details of the model for the higher twist are given. Numerical results are presented in Sec.~\ref{section_numerical}. 
Finally, in Sec.~\ref{section_conclusions} we state our conclusions.


\section{Longitudinal structure function $F_L$}
\label{section_longitudinal}

\noindent
The longitudinal structure function $F_L(x,Q^2)$  corresponds to the
interaction of the longitudinally polarized virtual photon in the
deep inelastic lepton-nucleon scattering. In the low $x$ region it is
  dominated by the  gluon
density. The experimental determination of $F_L$ is rather challenging since it
requires a measurement of the  dependence  
of the DIS cross-section on $y$, for fixed values of $x$ and $Q^2$.  
This in turn requires performing measurements at varying centre-of-mass 
energies, see e.g. Ref. \cite{H1:2013ktq}. Unlike the $F_2$ structure function, 
where the experimental data are abundant, the number of $F_L$ data points 
is rather limited so far and measurement errors are  rather large.

In the `naive' quark-parton model the structure function $F_L(x,Q^2)$ vanishes, as a consequence of quarks having  spin 1/2.
More precisely, $F_L$ vanishes when the transverse momenta of the quarks are limited. At large $Q^2$, $F_L$  is proportional to
($\langle m_q^2\rangle + \langle \kappa_T^2\rangle)/Q^2$, where $m_q$ is the
quark mass and $\kappa_T$, the quark transverse momentum, is by definition limited in the naive parton model. 
This remains approximately valid in the leading logarithmic
 order in the collinear approximation.
At higher logarithmic  orders  the point-like
QCD interactions (gluon emissions) allow 
for the average transverse momenta $\langle\kappa_T\rangle$ 
to grow  with increasing $Q^2$.
Thus  at higher orders of perturbation theory $F_L(x,Q^2)$ acquires a leading twist contribution. 
  In the limit of small $x$,
the longitudinal structure function is driven by the gluons through 
the $g \rightarrow q \overline {q}$ transition and thus permits a direct
measurement of the gluon density in the nucleon.

In the limit $Q^2\rightarrow 0$ the structure function $F_L$ has to vanish
as $Q^4$ (for fixed $2p\cdot q$) which  eliminates potential singularities at $Q^2=0$ of the hadronic
tensor $W^{\mu\nu}$.
It  also reflects the vanishing of the cross section $\sigma_L\sim F_L/Q^2$ in the 
 real photoproduction limit.

The longitudinal structure function is theoretically fairly
well understood at high $Q^2$, thanks to the framework of perturbative QCD. 
On the contrary, very little is known
about its  extrapolation towards the region of low $Q^2$; 
it is possible that it contains large contributions from higher twists there.

\newpage
\section{The $k_T$ factorization}
\label{section_kT}

\noindent

\begin{figure}[h]
\centering
\includegraphics[width=0.35\textwidth]{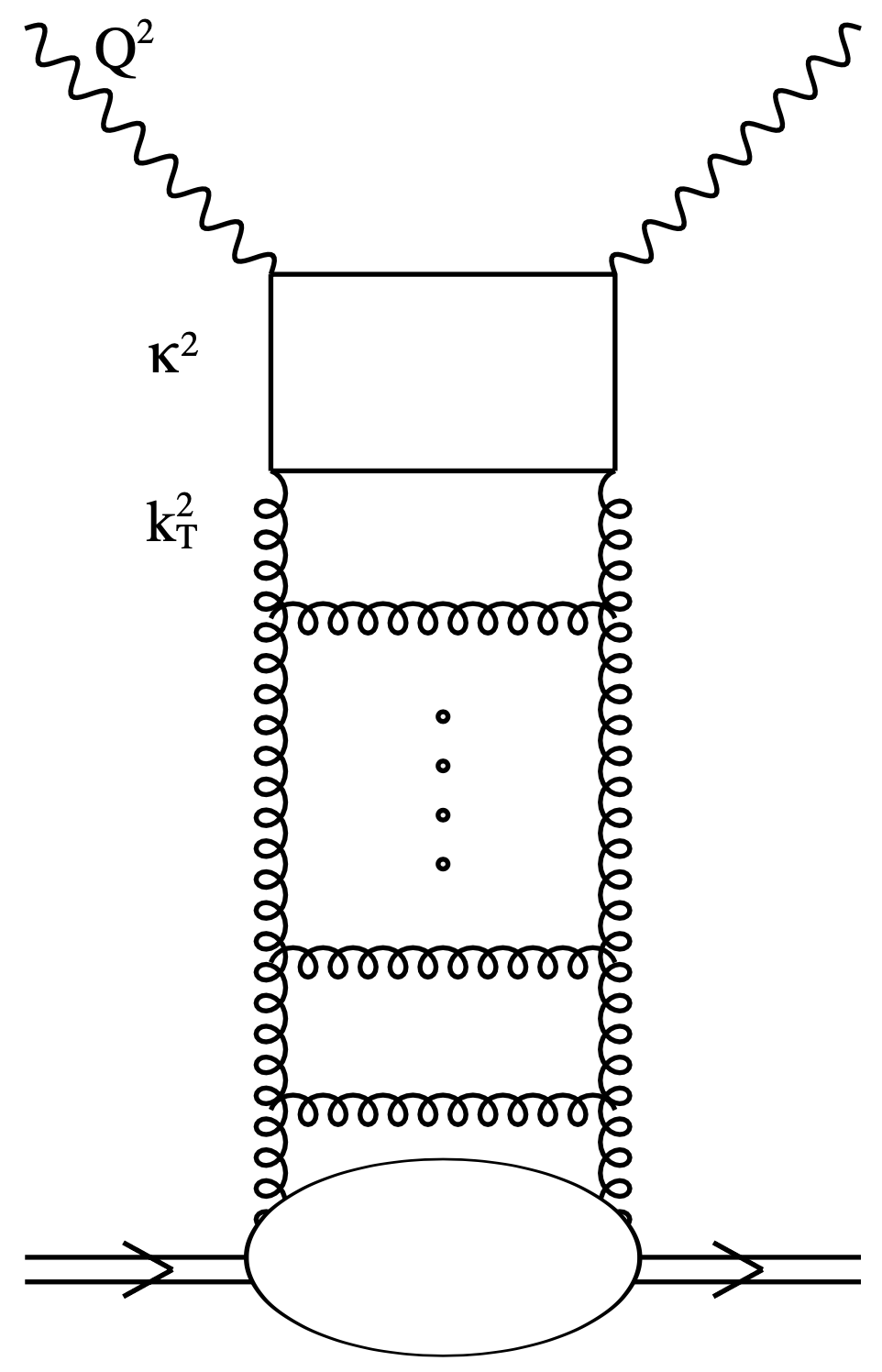}
\caption{Diagrammatic representation of the photon-gluon fusion mechanism and of the $k_T$-factorization formula. Symbols $\tdm{\rm\kappa^2}$ and $\tdm{\rm k_T^2}$ denote the transverse quark momentum squared and gluon transverse momentum squared, respectively.}
\label{fig:kt}
\end{figure}

Structure functions in the limit of small $x$ and at large $Q^2$ 
can be evaluated using the $k_T$-factorization theorem \cite{Catani:1990xk,Catani:1990eg,Collins:1991ty,Catani:1994sq,Ciafaloni:1995np,Levin:1990ef,Blumlein:1993ec}.
The basic process  which we consider is
 photon - gluon fusion, $\gamma^*+g\rightarrow q+\bar{q}$ , and 
the corresponding diagram  in the lowest order of perturbation theory 
is shown in Fig.~\ref{fig:kt}.
The distinctive feature of the $k_T$-factorization formula 
is the fact that the gluon transverse  momentum, here ${k_T}$, is taken
into account, unlike in the collinear factorization framework, 
where the integrated parton density depends only on the fraction 
of the longitudinal nucleon momentum  and the scale.\\

The photon-gluon  fusion  off-shell amplitude 
is known  up to NLO accuracy \cite{Bartels:2001mv,Bartels:2002uz,Balitsky:2012bs}. 
In our model, we use 
 the LO expression for  that amplitude but with the additional corrections 
stemming from the exact kinematics, which effectively takes 
into account a part of the important higher order corrections \cite{Bialas:2000xs,Bialas:2001ks}.

The expression for the longitudinal structure function 
can be written in the following form
\begin{equation} 
F_L(x,Q^2)=2 {\frac{Q^4} {\pi^2}} \sum _qe_q^2 I_q(x,Q^2) \;,
\label{eq:flsum} 
\end{equation}
where  $I_q$ is defined by
\begin{equation} 
I_q(x,Q^2)=\int {\frac{dk_T^2}{k_T^4}} \int_0^1 d\beta \int d^2\tdm{\kappa_T^{\prime}}  \, 
\alpha_s \,  \beta^2 (1-\beta)^2 {\frac{1} {2}}\left({\frac{1} {D_{1q}}} - {\frac{1}{D_{2q}}}\right)
^2f(x_g,k_T^2)  \; .
\label{eq:flint}
\end{equation}
The denominators $D_{1q},D_{2q}$ are defined as
$$ D_{1q}=\kappa_T^2 + \beta (1-\beta) Q^2 + m_q^2  \, ,$$
\begin{equation}
D_{2q}=(\tdm{\kappa_T} - \tdm{k_T})^2 + \beta (1-\beta) Q^2 + m_q^2\;.
\label{eq:denominator} 
\end{equation}
In  Eq.~(\ref{eq:flsum}) the sum runs over all the active quark 
flavours $q$, here u, d, s, c. The $\alpha_s$ is the strong coupling. 
The transverse momenta of the quark and gluon are denoted by 
$\tdm{\kappa_T }$  and $\tdm{k_T }$ respectively, see Fig.~\ref{fig:kt}. 
Let us define the shifted transverse momentum  $\tdm{\kappa_T^{\prime}}$ 
as follows
\begin{equation}
\tdm{\kappa_T^{\prime}} = \tdm{\kappa_T} - (1-\beta)\tdm{k_T } \; .
\label{kappap}
\end{equation}

The variable
$x_g$  is the fraction of the longitudinal momentum of the proton carried by the gluon which can be computed by taking into account the exact kinematics \cite{Askew:1992tw} 
\begin{equation}
x_g=x\left(1 + {\kappa_T^{\prime 2} + \frac{m_q^2} {\beta (1-\beta)Q^2}} + 
{\frac{k_T^2} {Q^2}}\right)  \; .
\label{yq}
\end{equation}
 The variable $\beta$ is the corresponding Sudakov
parameter appearing in the quark momentum decomposition into the basic
light-like four vectors,  $p'$ and $q'$ which are defined as
\begin{equation}
p'=p-{\frac{M^2 x } {Q^2}}q   \, , \;\;\;\;
q'=q+xp \; ,
\end{equation}
where $M$ is the target nucleon mass. 
We  decompose four-vector
$\kappa$  as 
\begin{equation}
\kappa  =x_qp'-\beta q'+\tdm{\kappa_T} \; , 
\label{decomp}
\end{equation}
where
\begin{equation}
x_q  =x\left(1+\frac {m_q^2+\kappa_T^2} {(1-\beta)Q^2}\right)  \; .
\end{equation}
\noindent
The variable $x_g$ in \eqref{yq} is the fraction of the longitudinal momentum 
of the proton carried by the gluon, which couples to the quarks 
in the `quark-box', see Fig.\ref{fig:kt}. In the leading order case 
this variable should be equal to Bjorken $x$. This is also the case 
in the dipole model formulation of the high energy scattering. 
Here however we include the exact kinematics,  which effectively 
makes this fraction  larger than   $x$. This has significant numerical effects, 
as demonstrated for example in Ref. \cite{Golec-Biernat:2009mod}.

The unintegrated gluon density $f(x_g,k_T^2)$ should in principle be 
obtained 
from an equation which resums small $x$ contributions, 
that is the BFKL equation. In Ref.\cite{Badelek:1996ap}
we used the approach where the unintegrated gluon density  was computed 
from the standard integrated density by taking 
a logarithmic derivative
\begin{equation}
f(y,k_T^2)=\left. y\frac{\partial g^{AP}(y,Q^2)}{\partial \ln Q^2}\right|_{Q^2=k_T^2},
\label{derivap}
\end{equation}
where $g^{AP}(y,Q^2)$ satisfies the conventional (LO or NLO) 
DGLAP equations.
In this approximation one neglects the higher order small $x$ resummation
effects in the
gluon density and therefore in the gluon anomalous dimension. 
This approximation will be used in the present update of the model\footnote{This result affects the gluon distribution only at very small values of $x \lesssim 10^{-4}$. }.
We shall also use the gluon density from the Kimber-Martin-Ryskin approach 
\cite{Kimber:1999xc,Kimber:2001sc}  which supplements the formula \eqref{derivap} with the Sudakov form-factor

\begin{equation}
f(y,k_T^2)=\left. y\frac{\partial \bigg( g^{AP}(y,Q^2) T(Q,k_{T}) \bigg) }{\partial \ln Q^2}\right|_{Q^2=k_T^2},
\label{eq:kmr}
\end{equation}
 defined as

\begin{equation}
\label{eq:3}
T(Q,k_{T}) =\exp\left\{
-\int_{k_T^2}^{Q^2} \frac{dp_T^2}{p_T^2}\int_0^{1-\Delta} dz z P_{gg}(z,p_T)
\right\} \, ,
\end{equation}\\
\noindent
with  the $P_{gg}$ the gluon-gluon DGLAP splitting functions.  
The cut-off $\Delta$ is necessary to re\-gu\-late the divergence at $z=1$. 
As discussed in Ref.\cite{Golec-Biernat:2018hqo},  different forms of the cut-off were considered 
in the literature: the strong ordering cut-off $\Delta=k_T/Q$ 
and the angular ordering cut-off $\Delta=k_T/(k_T+Q)$. 
We checked both choices, which in our calculation do not lead to any significant differences of the results.

The parameters $m_q$  in Eq.~\eqref{eq:denominator} are the masses of the quarks.
Since we are interested in the low $Q^2$ region of the $F_L$ structure function, 
the values of the masses are important. It should be noted that
the integrals $I_q$ defined by Eq.~\eqref{eq:flint}  are infrared finite
even if we set $m_q=0$.  However, the non-zero values of the quark masses 
are  necessary if formula \eqref{eq:flsum} is extrapolated down to $Q^2=0$, respecting the kinematic constraint $F_L\sim Q^4$. The non-zero quark masses then
play the role of the infrared regulator.

We shall consider two scenarios for the quark masses. In the first scenario we set the masses  equal to $m_q^2 \approx m_v^2/4$ where $m_v$ denotes 
the mass of the lightest vector meson which can be viewed as a component in the photon wave function.  
This non-perturbative Vector Meson Dominance approach can be argued within the  dipole model framework. In the latter one views the DIS process  
as the photon fluctuating into the $q\bar{q}$ pair, 
the color dipole which then scatters off the target through the exchange of gluons. This exchange can be related to the unintegrated gluon density function, $f(y,k_T^2)$.
We thus assume that the expressions for the structure function $F_L$
within photon-gluon fusion mechanism include also the virtual vector meson contributions.
This motivates the choice of the quark masses to be related 
to the vector meson masses.

In the second scenario, we will treat the quark masses as parameters, which can be tuned to obtain the best description of the data at small $Q^2$. As we shall see in more detail in Sec.~\ref{section_numerical} masses $m_q=140 \; $GeV for the u, d, s quarks provide an excellent description of the experimental data from JLab at low values of $Q^2$, as in the dipole model \cite{dipole}. 

Finally, it should be mentioned, that yet another process contributes to the $F_L$ structure function, which  originates from the virtual photon-quark interactions, with the emission of a gluon, $\gamma^* + q(\bar{q})\rightarrow q'(\bar{q}')+g$. This contribution is treated within the standard collinear approximation as in Ref.~ \cite{Altarelli:1978tq}. These two mechanisms will be referred to as gluon and quark contributions respectively.


\section{The model for higher twist contribution}
\label{section_model}

\noindent

\noindent
In this section 
we shall describe the 
model for the higher twist contribution to  
$F_L$ 
at low values of $Q^2$, following exactly Ref.\cite{Badelek:1996ap}.  As discussed in Sec.~\ref{section_longitudinal}  the structure function $F_L$
 contains significant higher twist contributions in the region of small 
$Q^2$ \cite{Miramontes:1989ni,SanchezGuillen:1990iq,Alekhin:1999iq}. Such higher twists vanish  
when $Q^2 \rightarrow \infty$.
For example, higher twists were analyzed  in the context of 
the dipole model, 
together with the saturation of the dipole cross section.
It was demonstrated that this framework  predicts large contributions from the higher twists to the longitudinal structure function \cite{Bartels:2000hv,Bartels:2009tu,Motyka:2017xgk}.

The  main idea of our approach \cite{Badelek:1996ap} was to consider separately 
the regions of low and high transverse momenta of the quarks and the gluons
in the proton. Integration  over $\tdm{\kappa'_T}$ in the integral \eqref{eq:flint}
is divided into the region of low and high transverse momenta,
$0<\kappa'^2_T<\kappa'^2_{0T}$, 
and $\kappa'^2_T>\kappa'^2_{0T}$, respectively.
Here,  $\kappa'^2_{0T}$ is
an arbitrary  phenomenological cut-off parameter, chosen to be 
of the order of 1 GeV$^2$. We varied this parameter 
in the range $0.5-1.2$ GeV$^2$, and found the sensitivity of the results less than about 10\%.

The region of high transverse momenta is treated according 
to the expression from $k_T$-factorization, Eq.~\eqref{eq:flint}. 
In the low $\kappa_T^{\prime 2}$
region, which is likely to be dominated by the soft physics, we use
the `on-shell' approximation which corresponds to setting transverse momentum of the gluon to zero. This allows to cast the Eq.~\eqref{eq:flint} into the collinear form with the  gluon density $zg(z,Q^2)$ which is obtained from an unintegrated gluon density $f$, see Ref.~\cite{Badelek:1996ap} for details.  Next, we  make the substitution:
\begin{equation}
\alpha_s zg(z,Q^2) \rightarrow A\;,
\label{subst}
\end{equation}
where $A$ is a dimensionless parameter.
This leads to the following representation of the higher twist
contribution to $F_L$:
\begin{equation}
F_L^{HT}=2A\sum_q e_q^2 {\frac {Q^4} {\pi}}
\int_0^1 d\beta \beta^2 (1-\beta)^2 \int_0^{\kappa^{\prime 2}_{0T}}
 d\kappa_T^{\prime 2}\, \frac{\kappa_T^{\prime 2}}{D_q^4} \; ,
\label{flht}
\end{equation}
where the denominator $D_q$ is given by
\begin{equation}
D_q= \kappa_T^{ \prime 2} + \beta (1-\beta) Q^2 + m_q^2  \;.    \label{eq:dq0}
\end{equation}
The constant $A$ in Eq.~\eqref{subst} can be fixed  from the transverse structure function,  $F_T$. We 
assume that the non-perturbative contribution to $F_T$  also comes from the region of low values of 
$\kappa_T^{\prime 2}$ and is controlled by the same parameter $A$.
The term $F_L^{HT}$ does not depend on $x$  and thus can be interpreted as representing the contribution
of soft Pomeron exchange with intercept 1.

It should be noted that $F_L^{HT}$ given by equation (\ref{flht})
 vanishes as $1/Q^2$ in the high $Q^2$
limit (modulo logarithmically varying factors).  We call it
therefore a `higher twist contribution'.  Observe that this term will also respect the kinematic
constraint $F_L \sim Q^4$ in the limit $Q^2 \rightarrow 0$.

In order to find the parameter $A$ one needs to consider the transverse structure function $F_T$ and perform the same approximations on it as above. 
In the on-shell case, that is when the gluon transverse momentum is small, the corresponding formula which describes the photon - gluon contribution to the structure function $F_T$ is
\begin{eqnarray}
F_T(x,Q^2)&=&2\sum_qe_q^2 \frac{Q^2} {4 \pi} \alpha_s
\int_0^1 d\beta \int d\kappa_T^{\prime 2} \,x_g\,
g(x_g,Q^2) \times \nonumber \\
&& \times\left[\frac{\beta^2 + (1 -
\beta)^2}{2} \left ({\frac{1}{D_q^2}}- \frac{2 \kappa_T^{\prime 2}}{D_q^3} +\frac{2\kappa_T^{2}\kappa_T^{\prime 2}}{D_q^4}\right) 
+\frac{m_q^2 \kappa_T^{\prime 2}}{D_q^4} \right]\;.
\label{ft}
\end{eqnarray}
The soft term in $F_T$ is then obtained from this expression  by integrating over 
the low $\kappa_T^{\prime 2}$ region
($\kappa_T^{\prime 2}<\kappa_{0T}^{\prime 2}$).
Finally, we perform the same substitution as in Eq.~\eqref{subst} and
 we identify this
 soft part of 
 $F_T$ with the `background' 
 term $F_2^{Bg}$ of Ref.\cite{Askew:1992tw}, i.e.
\begin{equation} 
F_2^{Bg}=A\times\frac{\sum_qe_q^2}{\pi} \int _0^{\infty} dt
 \int _0^{\kappa_{0T}^{\prime 2}}d\kappa_T^{\prime 2} 
\left[\frac{1}{2} \left (\frac{1}{D_q^2}- \frac{2 \kappa_T^{\prime 2}}{D_q^3} +\frac{2\kappa_T^{2}\kappa_T^{\prime 2}}{D_q^4}\right) 
+\frac{m_q^2 \kappa_T^{\prime 2}}{D_q^4} \right]    \; , 
\label{ftht}
\end{equation}
where now
\begin{equation}
D_q=m_q^2 + \kappa_T^{\prime 2} + t \;.
\label{dqft}
\end{equation}
A possible weak $x$ dependence of $F_2^{Bg}$ is neglected and we set
$F_2^{Bg}=0.4$, see Ref.\cite{Askew:1992tw}. \\
\noindent
The complete structure function $F_L$ is represented as
$F_L=F_L^{HT}+F_L^{LT}$
where $F_L^{LT}$ is calculated from Eqs~\eqref{eq:flsum} and \eqref{eq:flint} and $F_L^{HT}$ from Eq. \eqref{flht}.


\section{Numerical results}
\label{section_numerical}

\noindent
In this section we present  numerical results for the 
longitudinal structure function $F_L$ and compare them with
the experimental data.

\begin{figure}[ht]
\centering
\includegraphics[width=0.75\textwidth]{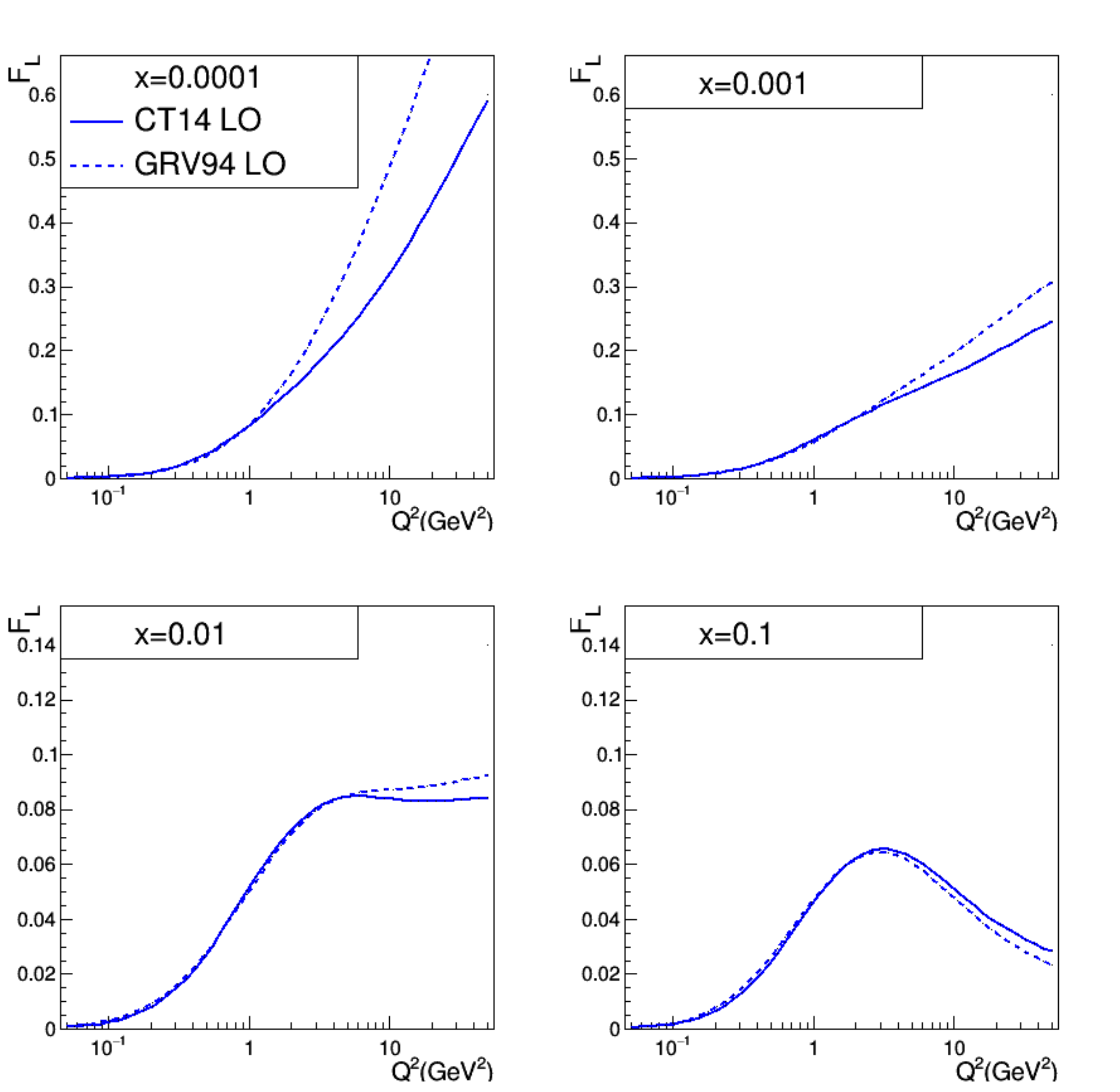}
\caption{The results from the model for $F_L$ for two different PDFs, CT14LO \cite{Buckley:2014ana} and GRV94LO \cite{GRV94}, as functions of $Q^2$, in bins of $x$. Observe  different vertical scales in upper and lower panels.}
\label{fig:fl_vs_q2_bins_x}
\end{figure}

\begin{figure}[ht]
\centering
\includegraphics[width=0.75\textwidth]{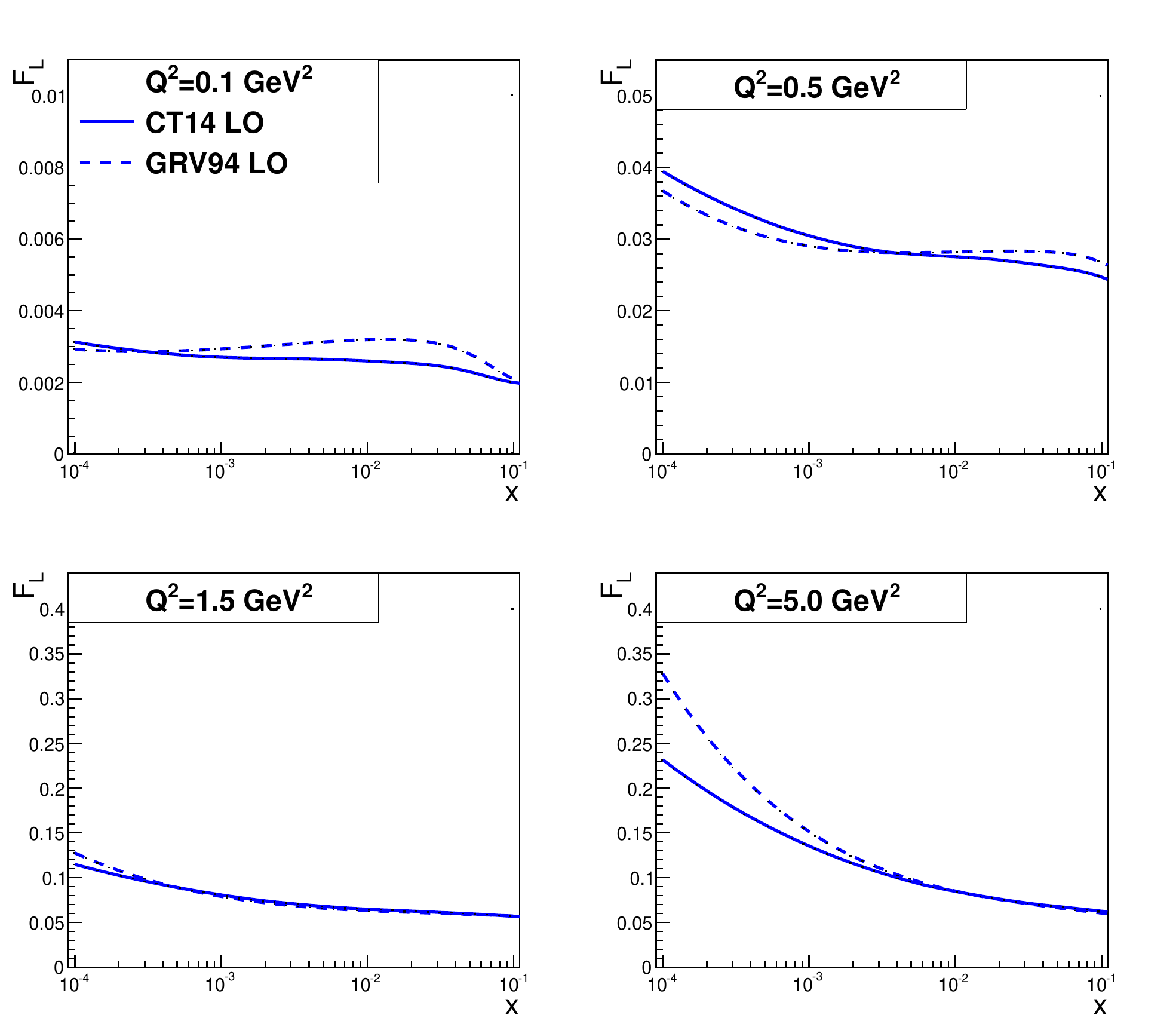}
\caption{$F_L$ as in Fig. \ref{fig:fl_vs_q2_bins_x} but as functions of $x$ and in bins of $Q^2$. Observe  different vertical scales in different panels.}
\label{fig:fl_vs_x_bins_q2}
\end{figure}

The calculated $F_L$ is shown in Figs 
\ref{fig:fl_vs_q2_bins_x} and \ref{fig:fl_vs_x_bins_q2} as function of $Q^2$ in bins
of $x$ and as function of $x$ in bins of $Q^2$, respectively. Two sets of the quark and gluon PDFs, both at the leading order accuracy, were used: the GRV94LO \cite{GRV94} (used previously in Ref. \cite{Badelek:1996ap}) and CT14LO \cite{Buckley:2014ana}. In the region of our interest, low $x$ and low $Q^2$ values the results practically do not depend on that choice.  We have also verified that changing from LO to NLO PDFs  has negligible impact on our results in the non-perturbative region. Therefore we stick to the LO choice of the PDFs. 
In what follows  the CT14LO parametrization\footnote{The more modern distributions, e.g. CT18 do not contain the LO PDFs.} will be used. 

Apart from the approach displayed in Eq. \eqref{derivap} we also used the unintegrated gluon density with the Sudakov form-factor, Eq.~\eqref{eq:kmr} and found negligible differences between these two approaches.
 Besides the dominant (at small $x$) photon-gluon fusion
$\gamma^* + g \rightarrow q + \bar{q}$ mechanism, 
we have also included  a 
contribution from quarks, $\gamma^* + q(\bar{q})  \rightarrow q'(\bar{q}')+g$ ,  taking into account threshold effects, for details see Ref.\cite{Badelek:1996ap}.  

In the calculations we took into account the  contributions from u, d, s and c quarks of masses 0.35, 0.35, 0.5 and 1.5 GeV respectively ($m_{q,h}$ masses). For comparison lower quark masses equal to 0.14 GeV for u, d, s quarks were also employed ($m_{q,l}$). The parameter $\kappa_{0T}^{\prime 2}$ was set to  0.8 GeV$^2$. This value was varied in the interval 0.5 -- 1.2 GeV$^2$ which resulted in changing the $F_L$ by at most  10\%.

In Ref.~\cite{Badelek:1996ap} we also  have performed  calculations within the on-shell approximation, which corresponded to setting the gluon transverse momentum to zero in the photon-gluon amplitude,  Eq.~\eqref{eq:flint} and restricting the integration over $k_T^2$ to $k_T^2 \ll Q^2$.  We found that the differences between off-shell and on-shell calculations are small, particularly in the low $Q^2$ region. Therefore in this update we only stick to the off-shell calculation.

\begin{figure}[ht]
\centering
\includegraphics[width=0.75\textwidth]{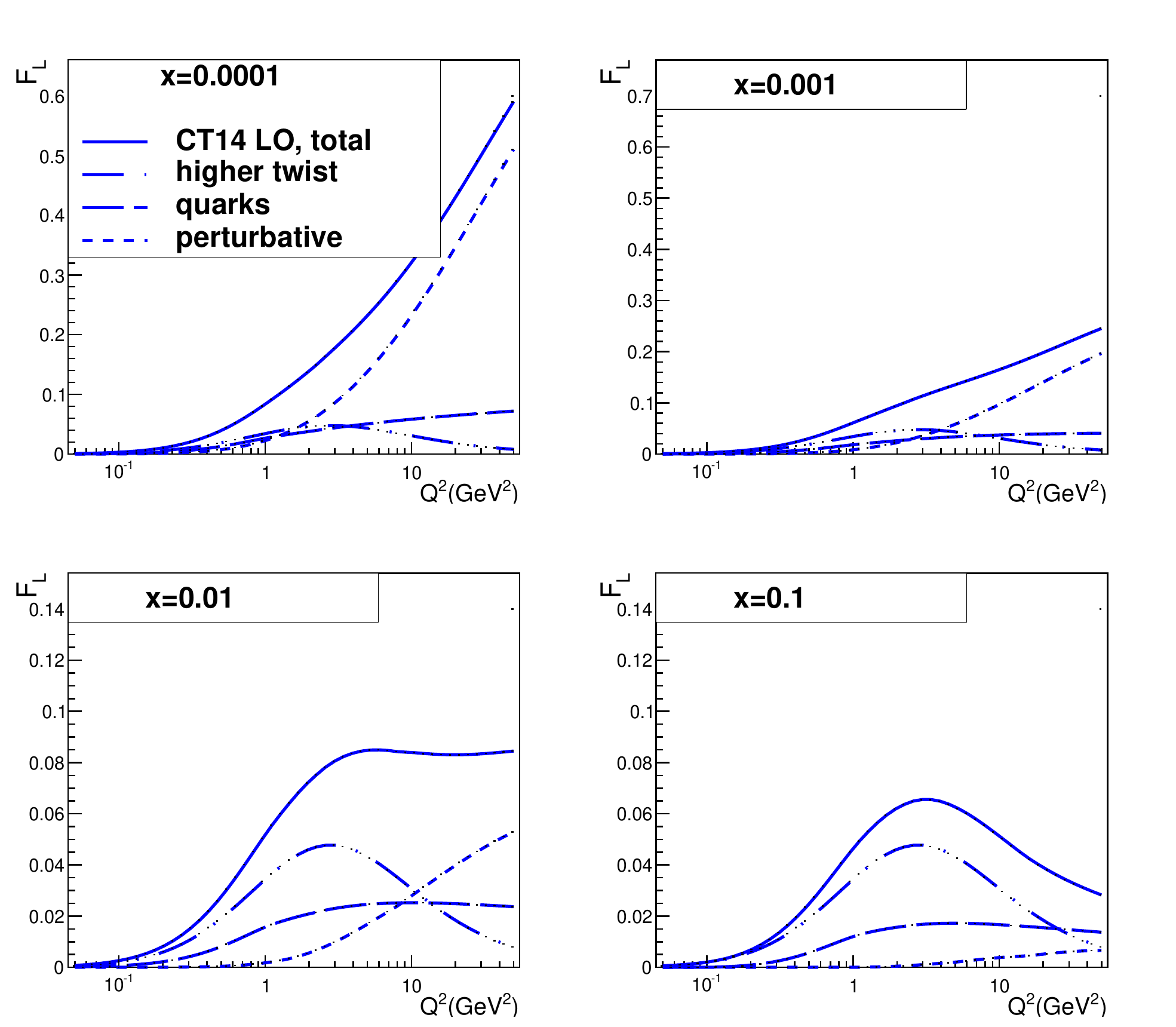}
\caption{Contributions building the model for $F_L$ as  functions of $Q^2$ and in bins of $x$. Observe different vertical scales in the upper and lower  panels.}
\label{fig:fl_vs_q2_bins_x_contrib_dis}
\end{figure}

\begin{figure}[ht]
\centering
\includegraphics[width=0.75\textwidth]{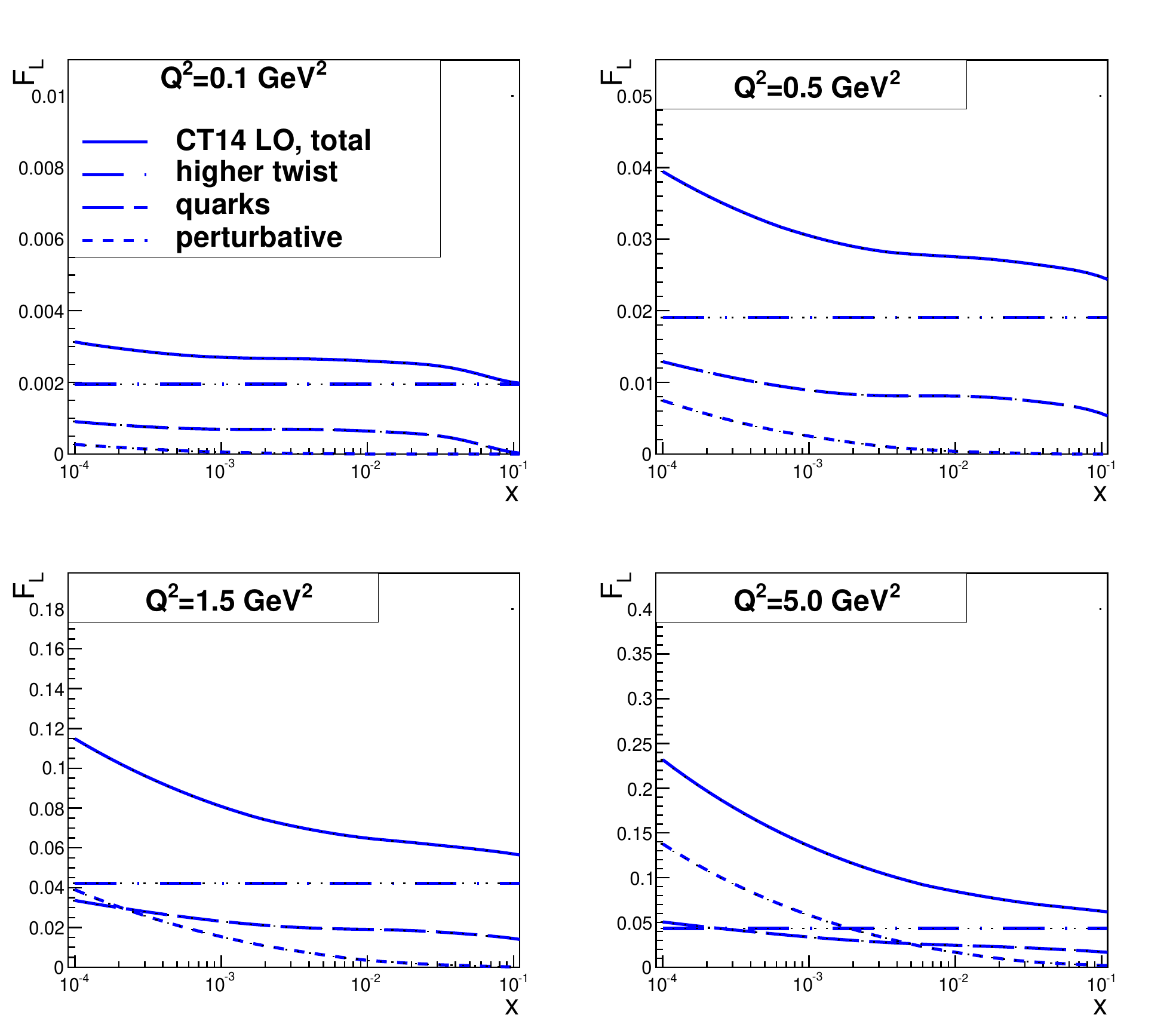}
\caption{Contributions to $F_L$ as in Fig. \ref{fig:fl_vs_q2_bins_x_contrib_dis} but as functions of $x$ and in bins of $Q^2$. Observe different vertical scales in different panels.}
\label{fig:fl_vs_x_bins_q2_contrib_dis}
\end{figure}

The (small) $x$-dependence is weak at low $Q^2$ and slightly growing
with decreasing $x$ while the $F_L$ at low $Q^2$ and low $x$ is very small, 
less than 0.005 (for $Q^2 \lesssim 0.1$ GeV$^2$) and strongly decreasing with decreasing $Q^2$. Contributions to $F_L$ from the quarks,
the perturbative part and the higher twist are illustrated in Figs
\ref{fig:fl_vs_q2_bins_x_contrib_dis} - \ref{fig:fl_vs_x_bins_q2_contrib_dis} and clearly
visualize the interplay of different mechanisms in building the $F_L$
in our model.   We note that the perturbative mechanism contributes very little in the low $Q^2$ region.

\begin{figure}[ht]
\centering
\includegraphics[width=0.75\textwidth]{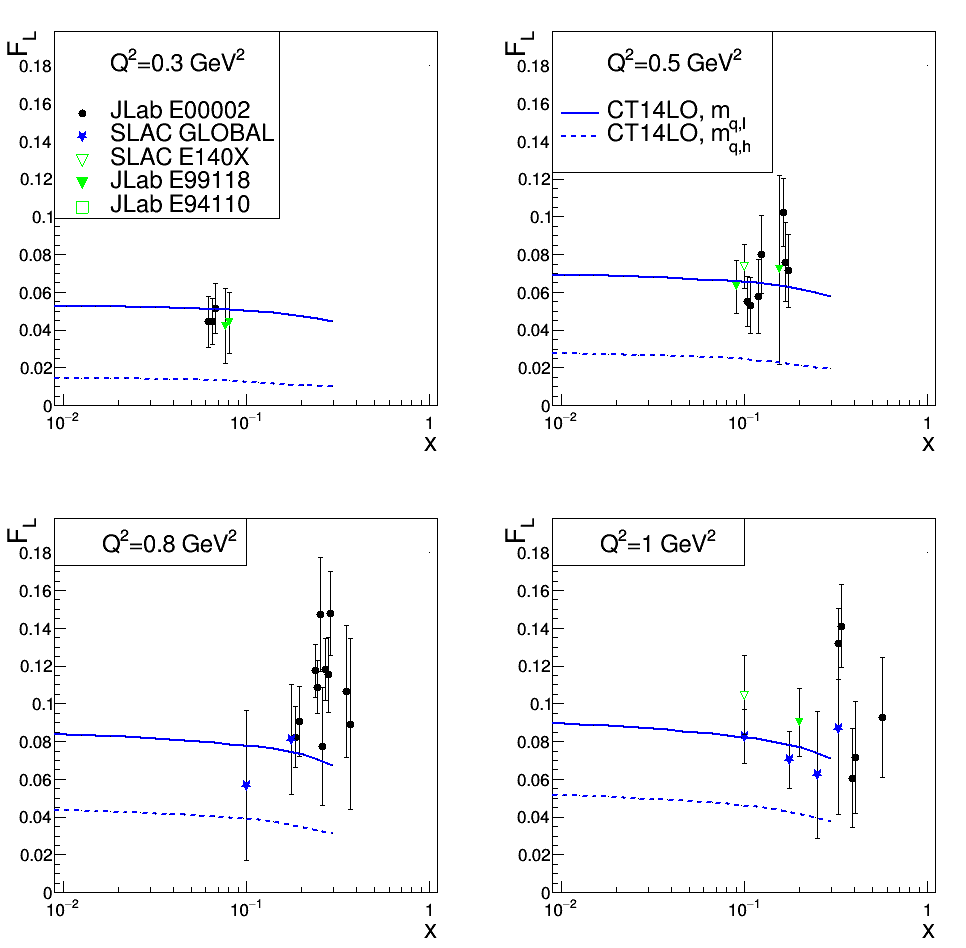}
\caption{Model calculations for $F_L$ for 
CT14LO PDFs \cite{Buckley:2014ana}, as functions of $x$ and in bins of $Q^2$, compared to 
the data of JLab \cite{Tvaskis:2016uxm,E99118,E94110} 
and SLAC \cite{E140X:1995ims,Whitlow:1991uw}.
The broken line marks calculations performed for quark masses $m_{q,h}$ while the continuous one - for lowered light quark masses, $m_{q,l}$ according to the dipole model \cite{dipole}.}
\label{fig:fl_vs_x_bins_q2_data}
\end{figure}

\begin{figure}[ht]
\centering
\includegraphics[width=0.75\textwidth]{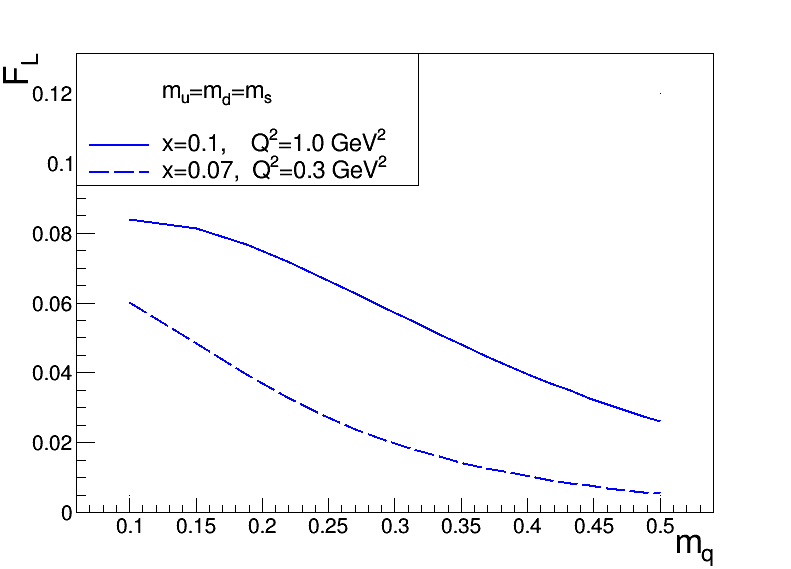}
\caption{Results of our model calculations for $F_L$ as function of quark masses, $m_q$ for two pairs of the $(x,Q^2)$ values.  Equal masses of u,d,s quarks are assumed.}
\label{fig:fl_vs_mq}
\end{figure}

In Fig.~\ref{fig:fl_vs_x_bins_q2_data} we compare the  calculations  with measurements. Unfortunately the latter
are very scarce in the non-perturbative region, see Fig.~1 
in Ref.\cite{Tvaskis:2016uxm} where a compilation of $F_L$ measurements is given. Only results from the JLab E99-118 \cite{E99118}, E94-110 \cite{E94110},
E00-002 \cite{Tvaskis:2016uxm}, SLAC E140X \cite{E140X:1995ims} 
and SLAC GLOBAL analysis \cite{Whitlow:1991uw} extend to the edge of the low $x$ non-perturbative region. There our model 
with standard light quark masses clearly underestimates the data
 (broken line in Fig.~\ref{fig:fl_vs_x_bins_q2_data}).
 On the other hand, results for the set of low  masses of the light quarks (u,d,s), 
 $m_{q,l}=0.14$ GeV which were used in the dipole model calculations \cite{dipole} seem to reproduce the measurements well (continuous line in Fig.~\ref{fig:fl_vs_x_bins_q2_data}). 
The strong  dependence of the results on the assumed quark
masses, is illustrated in Fig.~\ref{fig:fl_vs_mq} for two kinematic $(x,Q^2)$ points.

\begin{figure}[ht]
\centering
\includegraphics[width=0.75\textwidth]{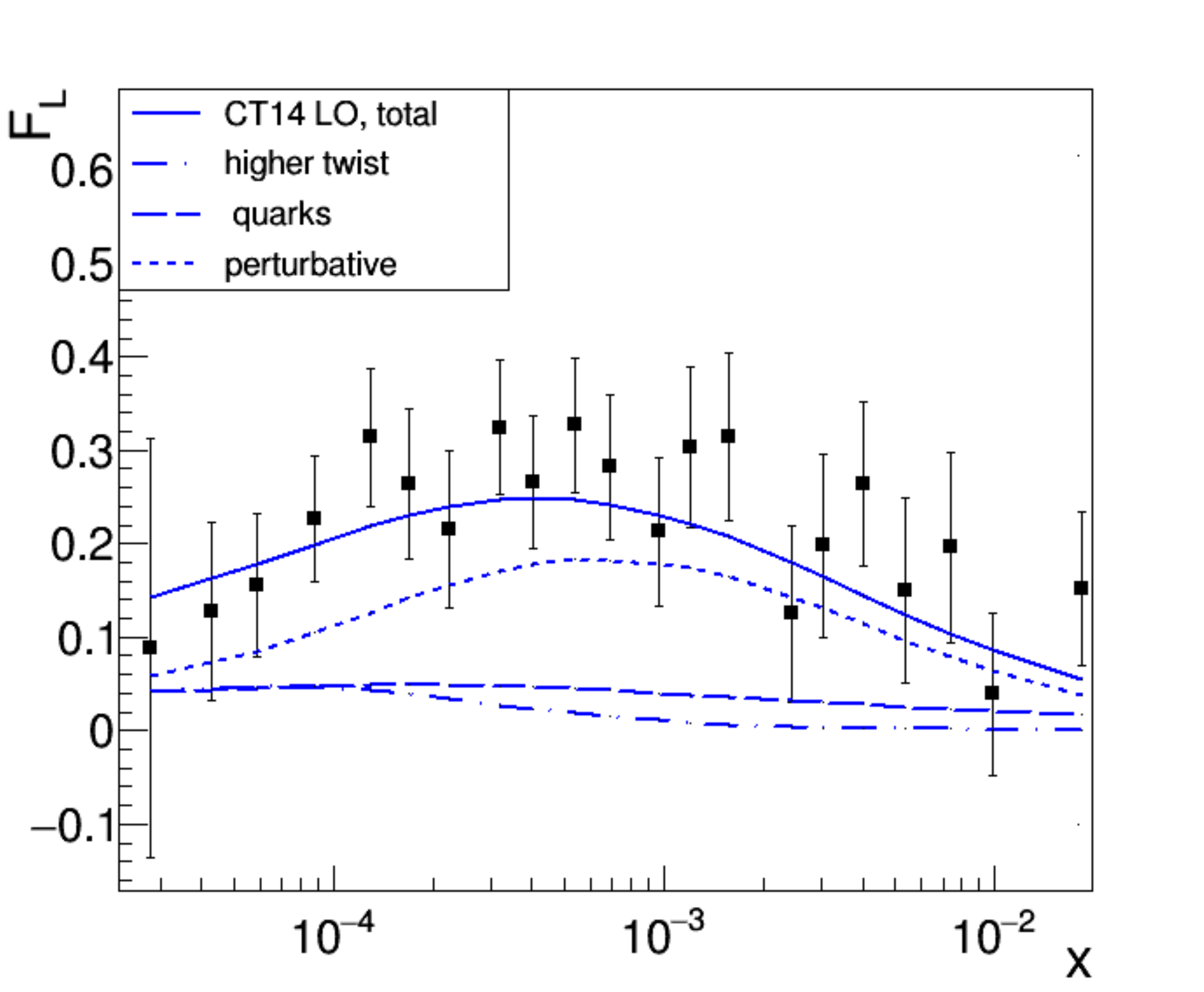}
\caption{Results from the  $F_L$ model with  CT14LO PDFs \cite{Buckley:2014ana} compared to the HERA data from H1 experiment \cite{H1:2013ktq}. Each data point corresponds to a different $Q^2$; the lowest $x$ point has lowest $Q^2$ equal to 1.5 GeV$^2$. Contributions to the model are also marked.}
\label{fig:hera}
\end{figure}

Finally, the H1 Collaboration measurements of $F_L$ in the perturbative region
at HERA, \cite{H1:2013ktq} are very well reproduced, see Fig.~\ref{fig:hera}.

\section{Conclusions}
\label{section_conclusions}

In this paper a parametrization of the proton longitudinal structure
function $F_L$ at low $Q^2$ and low $x$ is presented. It is a revisit and
update of the parametrization modelled in Ref.\cite{Badelek:1996ap} and based on the
photon-gluon fusion treated in the $k_T$ factorization,   suitably extrapolated 
to the region of low $Q^2$.
There is a severe need  to know $F_L$ in this region, down to $Q^2=0$, in the electroproduction data analysis 
 , e.g. in the QED radiative corrections.
Unfortunately  there are practically no $F_L$ measurements in this region. This leaves
the modelling constrained only at the upper end of the validity interval,
as (scarce) measurements are limited to  
$Q^2\gapproxeq$~0.1 GeV$^2$ and $x\gapproxeq$ 0.1, the latter 
being the  limit of applicability of our model.  

The model fulfills the kinematic constraint $F_L\sim Q^4$ at $Q^2\rightarrow 0$.
It also contains a higher twist term, 
which can be interpreted as a soft Pomeron exchange.  This term comes from the low transverse momenta of the quarks in the quark box diagram. The coupling of the soft Pomeron to the external virtual photon is modeled by a constant, which is fixed from the non-perturbative term of the transverse structure function. Thus our model for the higher twist in principle does not have any free parameters.

 As in the previous version of the model we have used the prescription for the unintegrated gluon density as  the logarithmic derivative of the standard integrated gluon density.
The unintegrated gluon density which includes the Sudakov  form-factor was also employed as an alternative. Compared to the old version of the model, we have used
newer parton distributions. The updated parton distributions  did not affect the results in the regime of low $x$ and low $Q^2$ in any significant way.

The structure function $F_L$ turns out to be very small at $Q^2\lapproxeq$ 
0.1 GeV$^2$, essentially independent of $x$ at low $x$ and low $Q^2$ and 
practically insensitive to the input parton distributions and
perturbative accuracy (LO or NLO) of the PDFs. The model of $F_L$ underestimates the JLab and SLAC measurements, 
 performed  at low $Q^2$ and  $x\sim 0.1$ unless the light quark masses are lowered down to 0.14 GeV, as in the dipole model, which brings the calculations to a very good agreement with the data. These data are however very scarce. On the 
other hand the model reproduces well the H1 measurements taken 
in the perturbative region ($Q^2\gapproxeq$ 1.5 GeV$^2$).
In conclusion, we believe that the presented model will be useful at several stages of the data analysis,  e.g. evaluation of the QED radiative corrections at the Electron Ion Collider.


\section*{Acknowledgments}

We thank    Fredrick Olness and Wojciech S\l{}omi\'nski  for discussions; we express our gratitude to 
Eric Christy, Cynthia Keppel, Peter Monaghan, Ioana Niculescu,   and Katarzyna Wichmann for supplying us with and discussing the measurements.
AMS is  supported by the U.S. Department of Energy grant No. DE-SC-0002145 and in part by National Science Centre in Poland, grant 2019/33/B/ST2/02588.
The code for calculating $F_L(x,Q^2)$ is available upon request from the authors.


\end{document}